\documentclass[final,5p,times,twocolumn]{elsarticle}

\usepackage{amssymb}
\usepackage{graphicx}
\usepackage{dcolumn}
\usepackage{bm}
\usepackage{epsfig}
\usepackage{color}
\usepackage{booktabs}
\usepackage{graphics}
\usepackage{amsmath}

\journal{}

\begin{document}

\begin{frontmatter}



\title{Studying the elastic properties of nanocrystalline copper using a model of randomly packed uniform grains}


\author{Guo-Jie J. Gao\corref{cor}}
\ead{jasongao@me.es.osaka-u.ac.jp, jason.gao@aya.yale.edu}
\author{Yun-Jiang Wang, Shigenobu Ogata}

\cortext[cor]{Corresponding author. Tel.: +81 668506198; Fax.: +81 668506097.}

\address{Department of Mechanical Science and Bioengineering, Osaka University, Toyonaka, Osaka 560-8531, Japan}

\begin{abstract}
We develop a new Voronoi protocol, which is a space tessellation
method, to generate a fully dense (containing no voids)
model of nanocrystalline copper with precise grain size control; we also perform
uniaxial tensile tests using molecular dynamical (MD) simulations to
measure the elastic moduli of the grain boundary and the grain
interior components at $300~K$. We find that the grain boundary
deforms more locally compared with the grain core region under thermal
vibrations and is elastically less stiff than the core component at
finite temperature. The elastic modulus of the grain boundary is lower
than $30\%$ of that of the grain interior. Our results will aid in the
development of more accurate continuum models of nanocrystalline
metals.

\end{abstract}


\end{frontmatter}



\section{Introduction}
\label{introduction}

Solid metallic systems usually possess polycrystalline structures
composed of crystalline grains of different sizes and
orientations. Polycrystalline metals with a grain size smaller than
100 nanometers are called nanocrystalline metals. Especially, because
of their small grain size, the grain boundary component occupies a
substantial part in these materials. We can think of nanocrystalline
metals as composites of crystalline grain cores and grain boundaries.
A grain boundary is the interface of finite average thickness and
nonzero area between two neighbouring crystalline cores. Both the grain
core and the grain boundary components play important roles in
determining the bulk mechanical properties of nanocrystalline metals
\cite{gleiter98, gleiter00, meyers06}. In this study, we focus on the
elastic modulus of the grain boundary component of nanocrystalline
metals. Understanding the elastic behaviour of the grain boundary
region would be beneficial in evaluating the stress field around crack
tips and dislocations, and can help elucidate the effect of porosity
on the mechanical properties of nanocrystalline metals.

In the past twenty years, it has been widely accepted that the grain
boundary is about $70$ to $75\%$ as stiff as the grain core component
\cite{shen95, kim99}. However, this belief is based on numerical
studies of specific grain boundaries, such as the relatively stable
$\Sigma 5$ twist boundary, where atoms interact with one another via
simplified Lennard-Jones potential at zero temperature \cite{kluge90},
or on experimental studies assuming that the grain boundary component
behaves like amorphous alloys
\cite{shen95,wong94,masumoto75,polk76}. A thorough investigation from
a general atomic scale structure at finite temperature is still
lacking.

In general, mechanical properties of nanocrystalline metals are affected by both grain size $d$ \cite{schiotz03, yip04} and its dispersity $\sigma_s(d)$ \cite{ma02}. Unfortunately, the microscopic structure of nanocrystalline metals has not been fully characterized experimentally, and different fabrication processes alter the microstructure extensively. Typically, Voronoi construction with randomly chosen Voronoi seeds is a model generating nonuniform grains and resembling nanocrystalline metals made by inert gas condensation \cite{schiotz99}. Here, we use Voronoi seeds from random close packing (RCP) (by that, we focus on its randomness and packing density of about $0.64$, not its strict mathematical definition \cite{debenedetti00}) of identical spheres to construct a new model of fully dense nanocrystalline copper with uniform grains and a well-defined $d$, the diameter of a spherical grain approximating the polyhedral grain created by the algorithm, so that we can investigate the grain size dependence of the elastic modulus $E$ separately, while ignoring the effect of $\sigma_s(d)$, a setup still experimentally unattainable, and has not been achieved by previous simulation studies. Using randomly packed uniform grains makes sure our model is not weakened on the granular scale due to the introduction of equal-sized grains and their possible ordered arrangement, and enables us to focus on the sub-granular structure such as the grain boundary component. Knowing this is an ideal model, we carefully establish its validity by inspecting its microstructure such as grain boundary thickness, the overall elastic and plastic behaviors. The results of all these tests agree very well with what have been reported in simulations and experiments, and therefore we believe our model is adequate to analyze the elastic modulus of the grain boundary component, a quantity beyond the approach of simplified bicrystal model or full-sized experimental measurement.

To study the elastic behaviour, we performed 3D simulations of uniaxial tensile tests at
$300~K$, about $22\%$ of the melting temperature of copper. It is
essential that this temperature is low, so that our models do not
alter their structures dramatically during tensile tests; on the other
hand, it is high enough for us to observe the thermal effect on the
elastic moduli, and there exist many other experimental results at
similar temperatures for comparison. We focused on systems with $d$
smaller than $25~nm$. Larger than this value, the grain core component
occupies the major part of the system and the elastic contribution of
the grain boundary component is negligible \cite{shen95,hytch03}.

\section{Nanocrystalline Model of Randomly Packed Uniform Grains and Molecular Dynamical Simulations}

The uniaxial tensile test has been considered the most direct way of
determining the mechanical properties of a material
\cite{legros00}. We implemented a new Voronoi-like algorithm to
generate the polycrystalline initial configurations for molecular
dynamics (MD) tensile tests. Unlike the conventional Voronoi
algorithm, where positions of Voronoi seeds are randomly chosen,
leaving $d$ ill-defined due to its high dispersity, particularly when
the number of grains is small, the positions of seeds in our algorithm
are mapped from center positions of 3D random close packings (RCPs) of monodisperse hard spheres, whose volume fraction is close
to the random close packing density, $\approx0.64$ \cite{berryman83}. Because
the average distance between any pair of Voronoi seeds is nearly a
constant in this algorithm, it allows us to generate almost
evenly-sized Voronoi grains and control the dispersity of $d$
accurately.

\subsection{Random Close Packing Finder of Identical Spheres}

\begin{figure}
\centerline{\includegraphics[width=0.5\textwidth]{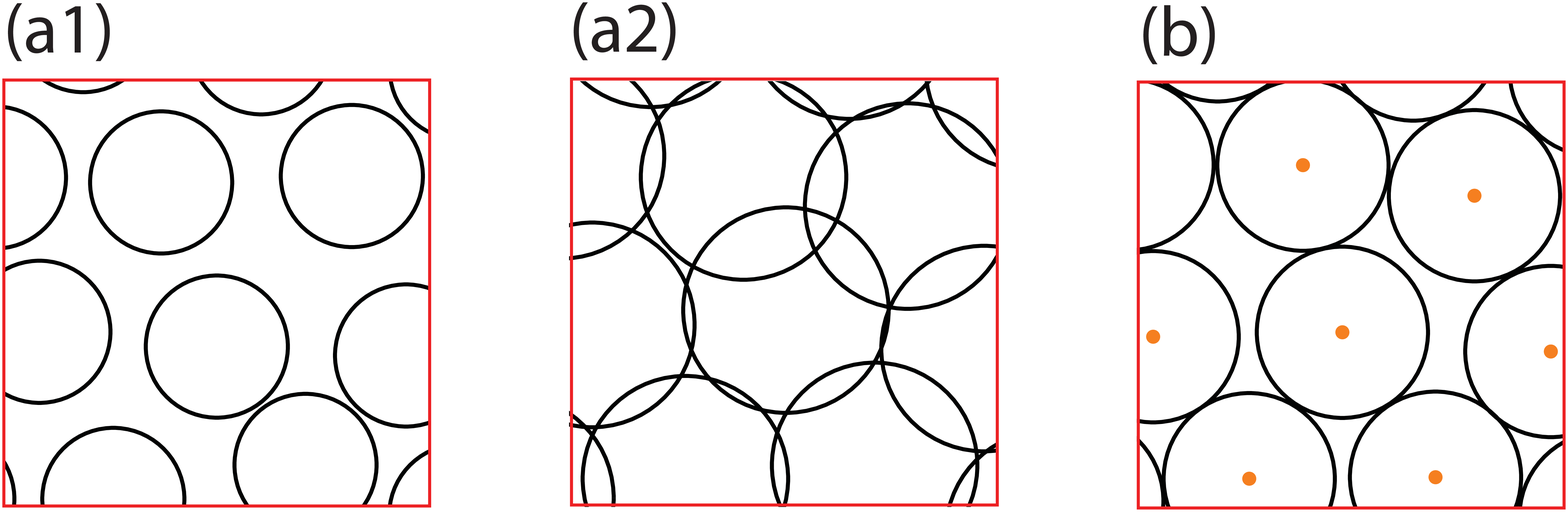}}
\caption{\label{fig:packing_algorithm} Algorithm for random close packings of identical spherical particles shown schematically in 2D. Intermediate states without overlap (a1) will grow, and overlapped intermediate states (a2) will shrink until a state (b) is reached. During the packing process, particles are allowed to move freely.}
\end{figure}

First we generate random close packings of identical spheres under
periodic boundary conditions. The algorithm begins with a non-overlapped random collection of spheres in a unit cell whose packing density is at least two orders of magnitude lower than the 3D random close packing density, $\approx0.64$. The interaction between spheres can be described by the soft, repulsive potential
\begin{equation} \label{spring_potential}
\frac{{{\bf V}_p (r_{ij} )}}{\epsilon } = \left\{ {\begin{array}{*{20}c}
   {\frac{1}{2}(1 - \frac{{r_{ij} }}{{R_{ij} }})^2 ,r_{ij}  < R_{ij} }  \\
   {0\begin{array}{*{20}c}
   {} & {} & {} & {}  \\
\end{array},r_{ij}  \ge R_{ij} }  \\
\end{array}} \right.,
\end{equation}
where $\epsilon$ is the characteristic energy scale, $r_{ij}$ is the distance between the $i$th sphere of diameter $R_{i}$ and the $j$th sphere of diameter $R_{j}$, and $R_{ij}=R_{i}+R_{j}$. The average potential energy of the system is given by the expression
\begin{equation} \label{total_potential}
{\bf V}_{tot}  = \frac{1}{n}\sum\limits_{i > j} {\frac{{{\bf V}_p (r_{ij} )}}{\epsilon }},
\end{equation}
where $n$ is the total number of spherical grains in the system.
After an initial configuration is generated, we grow (shrink) each sphere by $\Delta R$ so that the volume fraction $\phi$ of the system increases (decreases) by $10^{-4}$, if ${\bf V}_{tot}$ is smaller (greater) than a minimal average potential energy ${\bf V}_{min}=10^{-16}$ given by the machine precision. Each change of $\phi$ is followed by energy minimization of ${\bf V}_{tot}$ done by the conjugate gradient (CG) method. During the CG relaxation process, spheres are allowed to move freely. $\Delta R/2$ instead of $\Delta R$ is applied when switching from compression to expansion of the system and vice versa. The procedure of energy minimization following volume perturbation is repeated until ${\bf V}_{min} < {\bf V}_{tot} < 2{\bf V}_{min}$ to create a packing, where there are only tiny overlaps between spheres and each sphere keeps force balance with all its contact neighbors. The algorithm is schematically shown in figure~\ref{fig:packing_algorithm} \cite{xu05, gao06}.

The algorithm runs at zero temperature and under zero gravity, so the system is very unlikely to be trapped by global minima such as the fcc or hcp configuration on the potential energy landscape. In some rare cases, the packing algorithm ends up with unstable or ordered configurations, which can be removed manually. The number of grains $n$ and volume fraction $\phi$ of the specific sphere packings used to obtain the Voronoi seeds in this study are $(n, \phi)$ = $(8, 0.64)$, $(27, 0.63)$, and $(512, 0.64)$, respectively. All these sphere packings are randomly packed, and their $\phi$'s are basically the same as the RCP density. 

\subsection{Algorithm for Voronoi Tessellation}

\begin{figure}
\centerline{\includegraphics[width=0.4\textwidth]{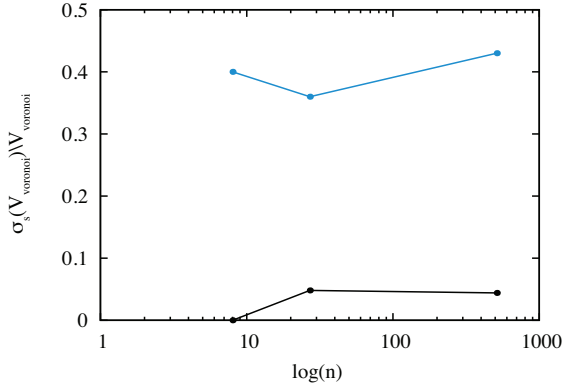}}
\caption{\label{fig:V_voronoi_dev_of_N} Normalized standard deviation $\sigma_s$ of Voronoi volume $V_{voronoi}$ for a single set of uniform Voronoi grains (black) and randomly chosen nonuniform grains (blue) in a unit cell. For all values of $n$ (8, 27, 512) used in this study, $\sigma_s(V_{voronoi})/ V_{voronoi}$ is smaller than $0.05$ for uniform Voronoi grains, but greater than $0.36$ for nonuniform grains. Volumes of grains are calculated using Voro++ \cite{rycroft09}.}
\end{figure}

\begin{figure}
\centerline{\includegraphics[width=0.48\textwidth]{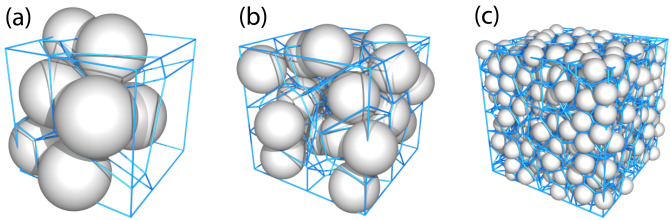}}
\caption{\label{fig:Voronoi_tessellation} Voronoi tessellations of randomly close packed identical spheres of (a) $n=8$, (b) $n=27$, and (c) $n=512$ used specifically in this study. Edges of Voronoi boundaries are shown by blue lines. Voronoi tessellations are visualized using Voro++ \cite{rycroft09}.}
\end{figure}

The centers of randomly close packed spheres are used as Voronoi seeds to
create Voronoi grains by the standard Voronoi tessellation: the $i$th
Voronoi grain contains points that are closer to the $i$th seed than
to any other seeds. The Voronoi boundary between the $i$th and $j$th spheres is determined by the relation
\begin{equation} \label{voronoi_boundary}
\frac{{r_{vi} }}{{R_i }} = \frac{{r_{vj} }}{{R_j }},
\end{equation}
where $r_{vi}$ is the distance between a point $v$ in the Voronoi boundary to the $i$th sphere of radius $R_i$. We call Voronoi seeds obtained this way `uniform Voronoi seeds' and the created Voronoi grains `uniform Voronoi grains'. In figure~\ref{fig:V_voronoi_dev_of_N}, it has been shown that the grain size dispersity $\sigma_s(d) \sim \sigma_s(V_{voronoi})$ of uniform Voronoi grains is very small  ($< 5\%$), compared with that of randomly chosen nonuniform Voronoi grains ($> 36\%$), and therefore influence of grain size dispersity is quantitatively negligible. Figure~\ref{fig:Voronoi_tessellation} shows the Voronoi tessellations of $N=8, 27, 512$ used specifically in this study. We fill each Voronoi grain with a face-centered cubic (fcc) randomly-oriented crystallite. Using
uniform Voronoi seeds guarantees that not a single Voronoi grain is
considerably larger or smaller than the average grain size, and each
grain is surrounded by grain boundaries with similar geometrical
features. In other words, the configuration can be treated as a
homogeneous medium down to the characteristic length scale $d$ in this
study. 

\begin{figure}
\centerline{\includegraphics[width=0.5\textwidth]{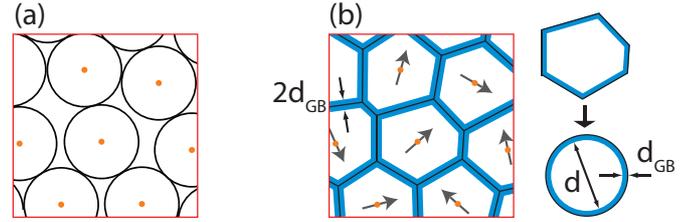}}
\caption{\label{fig:Voronoi_algorithm} A new
  Voronoi-like algorithm for generating nanocrystalline
  configurations: (a) First, create a packing of monodisperse
  spheres. Its schematic 2D section is shown here. Positions of the
  spheres (orange dots) are used as seeds in the standard Voronoi
  procedure to create uniform Voronoi grains. (b) Second, fill the uniform Voronoi grains with randomly-oriented fcc crystals (represented by
  black arrows), followed by MD relaxation at finite temperature. Each
  grain will then be surrounded by uniform grain boundaries (blue
  lines) with finite thickness. The size of the supercell has to be
  changed to get the desired grain size. We approximate these
  polyhedral Voronoi grains by spheres with an average diameter $d$
  and an average grain boundary thickness $d_{GB}$. The overall grain
  boundary thickness of the system is $2d_{GB}$.}
\end{figure}

Moreover, since the structure is randomly packed, it dismisses the concern that there may exist any artificial slip planes in the system caused by orderly arranged grains with a uniform grain size distribution, which may weaken the stiffness and strength of the system. After an initial configuration is constructed, under zero external loadings, we relax it at the temperature that is also used
for the subsequent tensile tests, until the MD extended hamiltonian
reaches a stable value. This relaxation process usually takes less
than $10^{-10}$ seconds. The new Voronoi-like algorithm is
schematically shown in figure~\ref{fig:Voronoi_algorithm}.

\subsection{Molecular Dynamical Simulation of Tensile Tests}

Using the Parallel Molecular Dynamics Stencil (PMDS) code
\cite{kimizuka03}, we performed three dimensional MD uniaxial tensile
tests, with periodic boundary conditions in all directions, on
nanocrystalline Cu at $300~K$. The normal stresses in the $x$ and $y$
directions perpendicular to the tensile direction $z$ are maintained
at zero, while the dimensions along these two directions are subjected
to free adjustment. The system exists in the isothermal-isostress
$N\sigma T$ ensemble \cite{martyna94}. Atoms interact with one another
via the embedded atom model (EAM) Mishin potential
\cite{mishin01,frolov09}. The strain-rate in the tensile direction is
$5 \times 10^{8} s^{-1}$. Young's modulus is strain-rate independent
though. We calculated the overall Young's modulus $E$ using data in
the strain ($\epsilon$) interval $\epsilon < 0.3\%$, which is clearly
within the linear stress($\sigma$)-strain($\epsilon$) region
\cite{schiotz99}.

\section{Validation of the Nanocrystalline Model Composed of Randomly Packed Uniform Grains}

\subsection{Geometrical Properties of the Grain Core and Grain Boundary Components}

\subsubsection{Identifying Atoms within the Grain Boundaries}

In general, atoms located between the interface of two grains are
treated as grain boundary atoms. To quantitatively measure the volume
fractions of the grain boundary and the grain core components, we use
two methods to sort out grain boundary atoms. The first method is
calculating coordination number (CN). Here, atom $i$ is a neighbour of
atom $j$ if the distance $r_{ij}$ between them is smaller than
$r_{cutoff}$, which is defined by the value of the first minimum of
the pair distribution function of the system. For Cu at room
temperature in our study, $r_{cutoff}=0.309~nm$. The total number of
neighbours of an atom is called the atom's coordination number. Atoms
of $CN<12$ are selected as grain boundary atoms and the rest are core
atoms. The second method is Common Neighbour Analysis (CNA). In this
method, we apply the same $r_{cutoff}$ for the conventional CNA
calculation using LAMMPS \cite{plimpton95, faken94, tsuzuki07} and
also try the adaptive CNA (a-CNA) calculation using OVITO
\cite{stukowski10, stukowski12}. Fcc and hexagonal close packed (hcp)
atoms are categorized as grain core atoms. Both CNA and a-CNA
calculations yield almost the same amounts of grain boundary and grain
core atoms.

\subsubsection{Measuring the Average Thickness of the Grain Boundaries}

Our Voronoi-like protocol generates grains shaped like convex
polyhedrons. We approximate these convex polyhedral grains by spheres
with an average diameter $d$ and an average volume $V$ satisfying the
relation
\begin{equation} \label{spherical_approx0}
V = \frac{{4\pi }}{3}\left( {\frac{d}{2}} \right)^3 = \frac{{N_{core}
    V_{core} }}{n},
\end{equation}
where $d$ is understood as $d_{CN}$ ($d_{CNA}$) if $N_{core}$, the
total number of the grain core atoms, is obtained by the CN (CNA)
calculation, $n$ is the total number of Voronoi grains, and $V_{core}$
is the average volume occupied by a single atom in an fcc or hcp unit
cell, which is $0.0118~nm^3$ for Cu at $300~K$.  Throughout our
investigation, $d_{CN}-d_{CNA}$ is always smaller than $0.4~nm$.

\begin{figure}
\centerline{\includegraphics[width=0.4\textwidth]{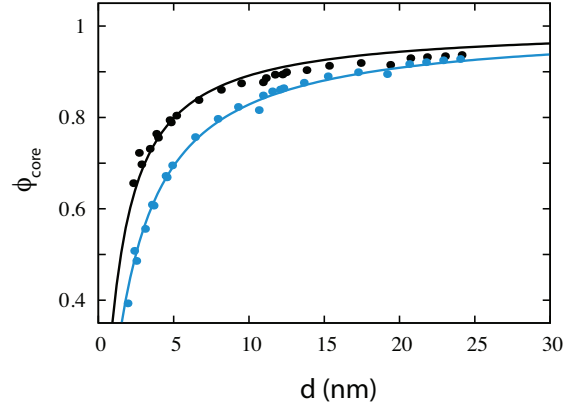}}
\caption{\label{fig:GB_thickness_fitting} Volume fraction
  of the grain core $\phi_{core}$ as a function of $d$
  ($d=d_{CN}$:black, $d=d_{CNA}$:blue) of nanocrystalline
  configurations generated by the new Voronoi-like algorithm (solid
  circles). $d_{CN}-d_{CNA}<0.4~nm$ through the whole range of $d$. We
  use $n=8$ for $d \ge 12~nm$, $n=27$ for $4~nm \le d \le 12~nm$, and
  $n=512$ for $d \le 4~nm$ to keep the normalized standard deviation
  $\sigma_{s}(d_{CN})/d_{CN} \le 0.1$. Solid lines are least
  square fits of grain boundary thickness $d_{GB}$ to
  Eq.~(\ref{spherical_approx2}). The estimated overall grain boundary
  thickness ($2d_{GB}$) is about $0.39~nm$ (CN calculation) and
  $0.64~nm$ (CNA calculation). Because $d_{GB}$ is near a constant,
  $\phi_{core}\to1$ at the coarse-grained limit ($d\to\infty$).}
\end{figure}

We can define the volume fraction of the grain core as
\begin{equation} \label{spherical_approx1}
\phi_{core} = nV/V_{system},
\end{equation} 
where $V_{system}$ is the volume of the simulation
supercell. Moreover, we assume that each approximated spherical grain
is covered with a grain boundary shell of average thickness
$d_{GB}$. The overall grain boundary thickness of the system is thus
$2d_{GB}$, because grain boundary is the space between two adjacent
grains (see figure~\ref{fig:Voronoi_algorithm}(b)). Following the above
assumptions, $\phi_{core}$ can also be given by the expression
\begin{equation}
\label{spherical_approx2}
\phi _{core} = \left( {\frac{d}{{d + 2d_{GB} }}} \right)^3.
\end{equation}
Varying the size of the simulation supercell, we can create a set of
$(d,\phi_{core})$, where $d$ and $\phi_{core}$ can be calculated using
equations (\ref{spherical_approx0}) and (\ref{spherical_approx1})
separately. Both CN and CNA calculations give similar $\phi_{core}$
when $d\ge 10~nm$. As a result, when greater than this system
size, $d_{CN}$ and $d_{CNA}$ basically can be used interchangeably and
labeled simply as $d$. By fitting the set to equation
(\ref{spherical_approx2}), we can estimate the grain boundary
thickness in our system as shown in
figure~\ref{fig:GB_thickness_fitting}. We found that the total fitted grain
boundary thickness of the system ($2d_{GB}$) is about $0.39~nm$ (CN
calculation) and $0.64~nm$ (CNA calculation), which are close to the
values ($\approx0.5~nm$) reported in other experimental and simulation
studies \cite{shen95,champion98,fultz00,pedersen09,stern95}. We also
directly calculated the grain boundary thickness by dividing the total
volume of the grain boundary component by the total interfacial area
between grains calculated by Voro++ \cite{rycroft09}. The thickness is
about $0.47~nm$ (CN calculation) and $0.53~nm$ (CNA calculation), as
the grain size $d$ is $25~nm$, where $E$ approaches the coarse-grained
limit.

\subsubsection{Geometrical Inhomogeneity of the Grain Boundary Component}

\begin{figure}
\centerline{\includegraphics[width=0.4\textwidth]{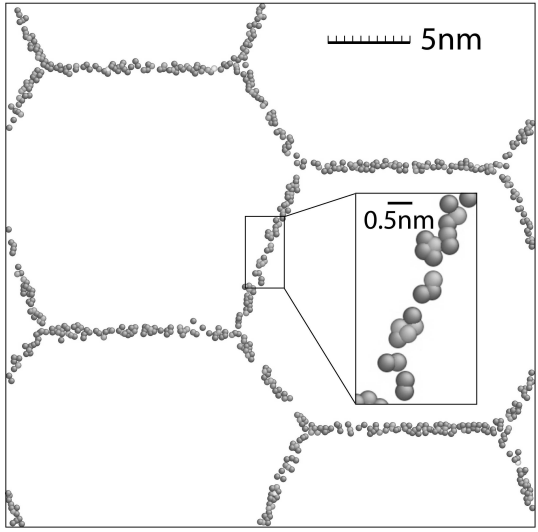}}
\caption{\label{fig:GB_sample} 3D snapshot of the grain boundary
  structure of a system of $d\approx21~nm$, where
  $(d_{CN},\phi_{core})=(20.7~nm,0.93)$ and
  $(d_{CNA},\phi_{core})=(20.6~nm,0.92)$. The dimension in the
  direction perpendicular to the figure plane is about one atom
  thick. We make all atoms of $CN=12$ invisible using Atomeye
  \cite{li03}. The inset is a blowup of the selected region. The grain
  boundary measures about two to three atoms thick with geometrical
  inhomogeneity. The configuration shown here may resemble an ordered packing visually, because the system contains only eight grains $(n = 8)$ in 3D, only four of which are shown on the 2D local profile. Refer to figure~\ref{fig:Voronoi_tessellation}(a) for its complete 3D Voronoi tessellation.}
\end{figure}

In figure~\ref{fig:GB_sample}, we show the grain boundary
structure. The grain boundary is about two to three atoms thick,
translating to $\approx0.5~nm$ for Cu at $300~K$. We can also observe
some geometrical inhomogeneity in the grain boundary region. This is
an intrinsic feature of the grain boundary, enhanced by thermal
fluctuations when the temperature is above zero. To analyze the
sparsity of grain boundary atoms due to this inhomogeneity, we
measured the average volume surrounding one atom in the grain
boundary, $V_{GB}$. The value is $0.01442\pm0.001~nm^3$ (CN
calculation) and $0.01234\pm0.0002~nm^3$ (CNA calculation), which are
$22\%$ and $5\%$ larger than $V_{core}$, respectively. Moreover, we
calculated the average CN of the whole system ($CN_{system}$) and of
the grain boundary ($CN_{GB}$). For $d_{CN}=12.2~nm$, we find they are
$11.9$ and $11.1$, respectively, which agrees with a recent
experimental study with a similar $d$ ($d=13~nm$, $CN_{system}=11.9$,
$CN_{GB}=11.4$) \cite{stern95}.

\subsection{Overall Elastic Response}

\begin{figure}
\centerline{\includegraphics[width=0.4\textwidth]{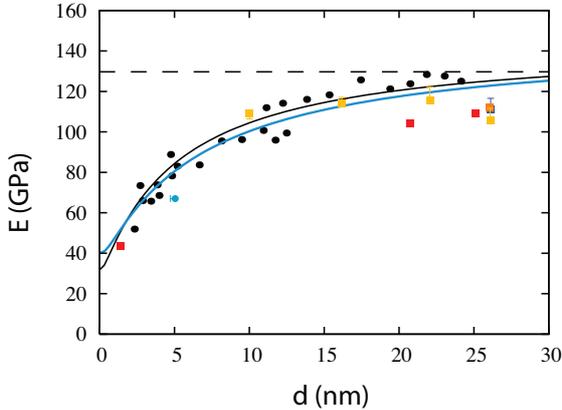}}
\caption{\label{fig:E_d} The overall Young's moduli $E$
  as a function of $d$. The black dots (original MD data), black (CN calculation, fitted), and blue (CNA
  calculation, fitted) lines are from this study; the cyan circle
  \cite{schiotz99} represents another simulation data; the yellow
  squares \cite{sanders97}, red squares \cite{sanders96,sharma03},
  orange square \cite{shen95}, and blue square \cite{legros00} show
  data from experimental studies; the dashed line represents the
  coarse-grained limit \cite{hertzberg96}. Color bars associated with
  the experimental data show projected values of the moduli at $300~K$
  by considering the temperature-dependence ($-40~MPa/K$) observed
  experimentally \cite{lebedev95}, if the original data were not
  obtained at this temperature. Color bar associated with the cyan
  simulation data shows the offset radius $d_{CN}$, because grain
  boundary thickness is not taken into account in that study.}
\end{figure}

The overall $E(d)$ is shown in figure~\ref{fig:E_d} against
experimental data \cite{sanders96,sharma03,legros00,sanders97,shen95},
and another MD simulation \cite{schiotz99} data. The experimental data
are obtained within a temperature range of $300~K$ to $450~K$,
however. To compare our simulation data with the experimental ones at
different temperatures meaningfully, we measured the
temperature-dependence of $E$ in our MD tensile tests. The obtained
value was $-62 \pm 7~MPa/K$, which is almost the same as what had been
reported in a previous MD study ($-60 \pm 18~MPa/K$) \cite{schiotz99},
but larger than one experimental value ($-40~MPa/K$) found in the same
material with a grain size of $200~nm$ \cite{lebedev95}. For a
temperature change of $150~K$ in simulation, Young's modulus will vary
by $10~GPa$, which is also the size of the error bar that should be
applied to the simulation data when we make a raw
comparison. Similarly, we can calculate the projected values of the
experimental data at $300~K$ if the temperature-dependence of
$-40~MPa/K$ is taken into account. The fitted $E(d)$ agrees with experimental data
extremely well, and approaches the coarse-grained limit $130~GPa$
\cite{hertzberg96} when $d$ is larger than $25~nm$.

\subsection{Flow Stress of Plastic Deformation}

\begin{figure}
\centerline{\includegraphics[width=0.4\textwidth]{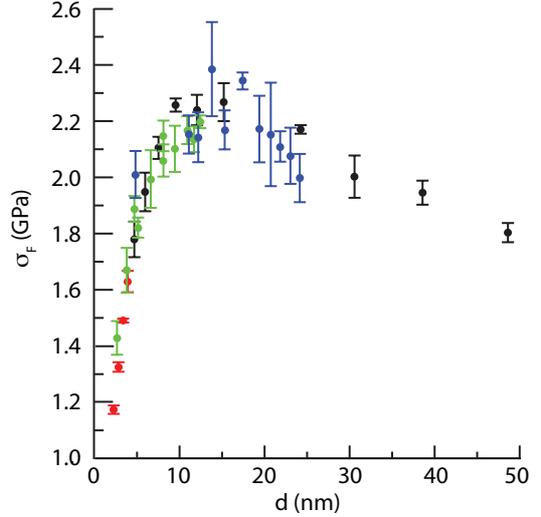}}
\caption{\label{fig:flow_stress_to_grain_size} Flow stress $\sigma_{F}$ as a function of uniform grain size $d$ compared with another simulation study using nonuniform Voronoi grains (black) \cite{schiotz03}. We use larger $n$ (8: blue, 27: green, 512: red) for smaller $d$ to prevent premature grain growth during relaxations and tensile tests due to small grain size. Both studies give similar plastic strength and Hall-Petch-like behavior.}
\end{figure}

Using the uniaxial tensile test at $300~K$ with a strain-rate $5 \times 10^{8} s^{-1}$, we also measured
the flow stress $\sigma_{F}$ taken as the average stress ($\sigma$) in the strain
($\epsilon$) interval between $7\%$ and $10\%$ as a function of $d$. In figure~\ref{fig:flow_stress_to_grain_size}, we compare our simulation data of uniform grains with another simulation study, where a set of Voronoi seeds are randomly chosen to create nonuniform grains \cite{schiotz03}. Both studies show a transition from regular Hall-Petch to inverse Hall-Petch behavior around grain size $d\approx15~nm$. Again, using uniform grains does not introduce extra weakness to our system due to monodispersity of the grain size distribution.

We also verified the isotropy of the system. For $n=8$ (smallest number of grains used in this study) grains of $d\approx25~nm$, we measured the elastic moduli and flow stresses in three perpendicular directions; their variations were only $6\%$ and $8\%$, respectively. For $n=512$ grains, the system contains $n\times(n-1)/2=130816$ different grain boundaries, and we believe it behaves isotropically. To make sure there is minimal finite size effect, we measured the elastic moduli and flow stresses of $n=8$ and $n=27$ grains of $d\approx12~nm$, and $n=27$ and $n=512$ grains of $d\approx5~nm$. The relative variations in the elastic moduli and flow stresses in either case are smaller than $12\%$ and $3\%$, respectively. In summary, all of the above agreements in the geometric, elastic, and plastic properties, and isotropy and finite size tests indicate that our new Voronoi-like protocol generates reasonably nanocrystalline configurations resembling real structures closely.

\section{Analysis of the Elastic Moduli of the Grain Core and Grain Boundary Components}

We can express the overall Young's modulus $E$ in terms of
$\phi_{core}$, $E_{core}$ and $E_{GB}$ by rules of mixtures, where
$E_{core}$ and $E_{GB}$ are Young's moduli of the grain core component
and the grain boundary component, respectively. To estimate $E_{core}$
and $E_{GB}$, we fitted $E$ to the Voigt \cite{voigt10}, Reuss
\cite{reuss29}, and Hill \cite{hill52} models, representing the upper
bound of $E$, the lower bound of $E$, and the arithmetic mean of the
previous two models, respectively. The Hill model has been chosen for
practical reasons \cite{karato08}. For a given set of $E$ and
$\phi_{core}$, $E_{core}$ and $E_{GB}$ become parameters to be
determined by numerical fitting. We tried both the $2-norm$ and
$\infty-norm$. The results are listed in Table~\ref{tab:table}.

\begin{table}
\centering
\caption{\label{tab:table} Young's moduli $(E_{core}, E_{GB},
  E_{GB}/E_{core})$}
\resizebox{\columnwidth}{!}{
\begin{tabular}{rrrr}
\toprule
(Unit: $GPa$) & Voigt & Hill & Reuss \\
\midrule
($2-norm$)\\
CN~ & $(138,-125, <0)$ & $(167,10, 0.06)$ & $(177,25, 0.14)$ \\ 
CNA & $(130,-8, <0)$ & $(158,20, 0.13)$ & $(153,40, 0.26)$ \\
\midrule
($\infty-norm$)\\
CN~ & $(136,-130, <0)$ & $(124,26, 0.21)$ & $(145,32, 0.22)$ \\ 
CNA & $(127,-7, <0)$ & $(136,32, 0.24)$ & $(146,40, 0.27)$ \\
\bottomrule
\end{tabular}
}
\end{table}

\begin{figure}
\centerline{\includegraphics[width=0.4\textwidth]{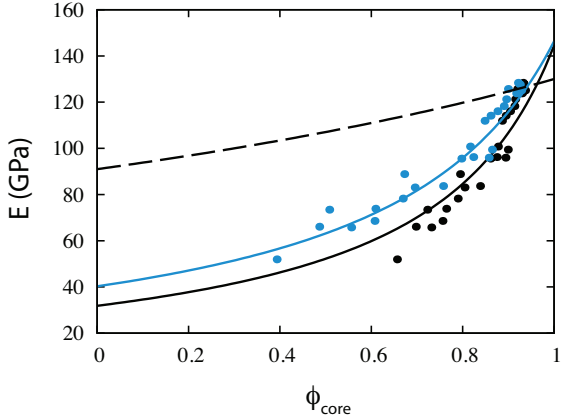}}
\caption{\label{fig:reuss_fit} Numerical fits with
  $\infty-norm$ (solid lines) of the overall Young's moduli $E$ of
  $\phi_{core}$ (solid circles) determined by CN (black) and CNA
  (blue) calculations to the Reuss model. A hypothetical Reuss curve \cite{reuss29} assuming $E_{core}=130~GPa$
  and $E_{GB}=0.7E_{core}$ (dashed line) is also shown to emphasize the disagreement between the hypothesis and the simulation data of this study.}
\end{figure}

The Voigt model gives unphysical negative values of $E_{GB}$ in all
four cases. Under both definitions of grain boundary atoms, Hill and
Reuss models give meaningful numbers worth further
consideration. Theoretically, the Reuss model describes the effective
moduli of a solid suspension in a matrix of near zero shear modulus, a
quantity proportional to the elastic modulus for homogeneous isotropic
materials \cite{avseth05}. In Table~\ref{tab:table}, we found that all
ratios of $E_{GB}/E_{core}$ are smaller than $0.27$, a close scenario
to an ideal Reuss example, although the ratio is nonzero. In other
words, in our system, grain cores are embedded in a grain boundary
matrix with a stiffness of less than about $1/4$ of that of the core
part. This result reciprocally justifies the fact that among the three
proposed models, the Reuss model best captures the elastic response of
our system. Specifically, assuming a linear constitutive relation, the
Reuss model can be expressed as
\begin{equation} \label{reuss_model}
E = \frac{1}{{\phi _{core} /E_{core} + (1 - \phi _{core} )/E_{GB} }}.
\end{equation}
The Reuss fits are shown in figure~\ref{fig:reuss_fit}. We choose the
results of $\infty-norm$ over $2-norm$, because the $E_{core}$ values
are much closer to the coarse-grained limit, $130~GPa$
\cite{hertzberg96}.

Knowing $d_{GB}$, $E_{core}$, and $E_{GB}$, we can rewrite $E$ as a
function of only $d$ by combining equations (\ref{spherical_approx2})
and (\ref{reuss_model}). The fitted overall $E(d)$ is shown in figure~\ref{fig:E_d} compared with other
experimental and simulation data \cite{sanders96,sharma03,legros00,sanders97,shen95, schiotz99}.

\begin{figure}
\centerline{\includegraphics[width=0.4\textwidth]{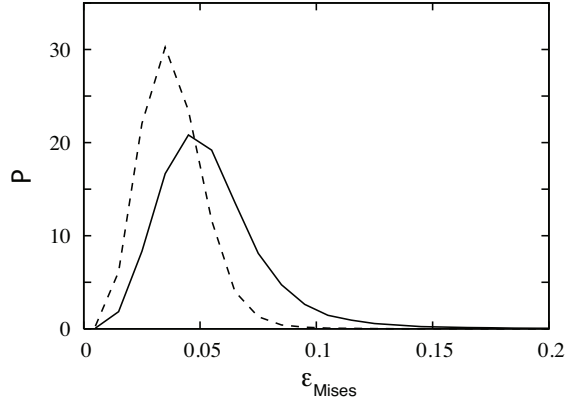}}
\caption{\label{fig:thermal_von_Mises_strain} Probability distribution
  of the least square von Mises shear strain \cite{shimizu07} of the
  grain core (dashed line) and the grain boundary (solid line)
  obtained by CN calculation under thermal fluctuation at
  $300~K$. $d\approx11~nm$, where
  $(d_{CN},\phi_{core})=(11.1~nm,0.89)$ and
  $(d_{CNA},\phi_{core})=(10.9~nm,0.85)$ in this system. The results
  are obtained by comparing two snapshots separated by $10~ps$. The
  grain boundary is subjected to more deformation even though there is
  no external loading.}
\end{figure}

The greater temperature-dependence in our system and the elastic softness of the grain boundary component may be because the
volume fraction of the grain boundary increases with decreasing $d$,
and the grain boundary is more thermally-sensitive than the grain
core, a fact due to the geometrical inhomogeneity in the grain
boundary (see the inset of figure~\ref{fig:GB_sample})
\cite{schiotz99}. We verify this by looking at atomic scale
deformation due to thermal fluctuations of both components at
$300~K$. In figure~\ref{fig:thermal_von_Mises_strain}, we show that,
locally, the grain boundary deforms more than the grain interior,
while both components are subjected to an equal amount of thermal
vibrations at the same temperature.

\section{Discussions and Conclusions}

In this study we investigated Young's moduli of the grain boundary and
the grain core components of fully dense nanocrystalline copper with
an average grain size $d \le 25~nm$ at $300~K$. We used an isotropic nanocrystalline model that includes not only the most stable grain
boundary, but also a general combination of varied grain boundaries
generated by our new Voronoi-like algorithm. The seeds of the new
Voronoi algorithm, used for creating monodisperse nanocrystalline
grains, are centers of randomly close packed identical spheres. The model composed of uniform Voronoi grains, although extremely ideal, faithfully reproduces the geometric, elastic, and plastic features of bulk nanocrystalline copper. We conducted MD uniaxial tensile tests to measure Young's modulus as a
function of $d$.

We found the following key results concerning the stiffness of
nanocrystalline Cu at finite temperature: 1) The grain boundary
component is less stiff than the grain interior; 2) At $300~K$,
Young's modulus of the grain boundary component is less than $30\%$ of
that of the grain interior, not $70\%$ as generally believed; 3) The
grain boundary component is more thermally-sensitive compared with the
grain core component, because the grain boundary shows more local
deformation compared with the core region under the same thermal
fluctuations.

It is important to emphasize here the distinct simplicity of our
approach to estimating Young's moduli at finite temperature. Starting
with a nanocrystalline model containing near-spherical monodisperse
grains with random crystal orientations, we obtain the grain boundary
thickness and its elastic modulus by numerical fits, without making
any assumptions about the elastic responses of the grain boundary. Due
to unavoidable inaccuracy of numerical fits, we focus on the ratio of
Young's moduli in the grain boundary and the core components, instead
of their absolute values. Through our analysis, we argue that the
grain boundary is elastically much softer, with less than $30\%$ of
the stiffness of the grain interior, at the studied temperature. This
striking finding could be explained by the fact that the somewhat
disordered grain boundary component is much more thermally sensitive,
and it has been verified by observing the structure change under
thermal fluctuations
(figure~\ref{fig:thermal_von_Mises_strain}). Because the CN of the grain
boundary component is only slightly smaller ($<7\%$ for
$d\approx12~nm$, for example) than that of the whole system and the
volume shared by a single grain boundary atom is merely $5\%$ to
$20\%$ bigger than that of a core atom, it is impressive that a small
disorder in structure can create a significant drop in stiffness.

Studies assuming Young's modulus of the grain boundary to be $70\%$ of
the grain interior value usually attribute the observed reduction of
the overall elastic modulus to porosity when the grain size is smaller
than $20~nm$ \cite{kim99}. The influence of porosity on the elastic
properties of nanocrystalline materials at finite temperature is
likely overestimated and should be further examined, since we have
shown here that the grain boundary is not as stiff as previously
believed.

Our results also directly help developing more advanced continuum
models. For example, in recent quasi-continuum simulations, it has
been shown that if the grain boundary has a very small elastic modulus
(17\% of the stiffness of the grain core), the shear stress field of
the edge dislocation shrinks with grain size \cite{shimokawa11}. Our
study about Young's modulus in the grain boundary region sheds new
light on this result.

\section*{Acknowledgments}

Financial support from grants for Scientific Research on the
Innovative Area of `Bulk Nanostructured Metals' (No. 22102003),
Scientific Research (A) (No. 23246025), Challenging Exploratory
Research (No. 22656030), Elements Strategy Initiative for Structural
Materials (ESISM), and JST under Collaborative Research Based on
Industrial Demand (Heterogeneous Structure Control) is gratefully
acknowledged. We also thank H. Kimizuka for advice about PMDS code,
T. Shimokawa and C. S. O'Hern for their insightful comments, and
I. Holca for proofreading the manuscript.





\bibliographystyle{elsarticle-num}
\bibliography{paper1}







\end{document}